\documentclass[journal]{IEEEtran}
\input epsf
\usepackage{graphicx}
\usepackage{bm}
\usepackage{mathabx}
\usepackage{cite}
\usepackage{tablefootnote}
\usepackage{float}
\usepackage{scalerel}
\usepackage{tikz}
\usetikzlibrary{svg.path}

\definecolor{orcidlogocol}{HTML}{A6CE39}
\tikzset{
  orcidlogo/.pic={
    \fill[orcidlogocol] svg{M256,128c0,70.7-57.3,128-128,128C57.3,256,0,198.7,0,128C0,57.3,57.3,0,128,0C198.7,0,256,57.3,256,128z};
    \fill[white] svg{M86.3,186.2H70.9V79.1h15.4v48.4V186.2z}
                 svg{M108.9,79.1h41.6c39.6,0,57,28.3,57,53.6c0,27.5-21.5,53.6-56.8,53.6h-41.8V79.1z M124.3,172.4h24.5c34.9,0,42.9-26.5,42.9-39.7c0-21.5-13.7-39.7-43.7-39.7h-23.7V172.4z}
                 svg{M88.7,56.8c0,5.5-4.5,10.1-10.1,10.1c-5.6,0-10.1-4.6-10.1-10.1c0-5.6,4.5-10.1,10.1-10.1C84.2,46.7,88.7,51.3,88.7,56.8z};
  }
}

\newcommand\orcidicon[1]{\href{https://orcid.org/#1}{\mbox{\scalerel*{
\begin{tikzpicture}[yscale=-1,transform shape]
\pic{orcidlogo};
\end{tikzpicture}
}{|}}}}

\usepackage{hyperref}
\begin{document}
\bstctlcite{IEEEexample:BSTcontrol}
\title{Improved Flexible Coaxial Ribbon Cable for High-Density Superconducting Arrays}

%\author{Jennifer Pearl Smith \orcidicon{0000-0002-0849-5867}, Benjamin A. Mazin  \orcidicon{0000-0003-0526-1114}, Alirio Boaventura, Kyle J. Thompson, Miguel Daal \orcidicon{0000-0002-1134-2116}}

\author{
    \IEEEauthorblockN{Jennifer Pearl Smith \orcidicon{0000-0002-0849-5867}\IEEEauthorrefmark{1}, Benjamin A. Mazin \orcidicon{0000-0003-0526-1114}\IEEEauthorrefmark{1}, Alirio Boaventura\IEEEauthorrefmark{2}, Kyle J. Thompson\IEEEauthorrefmark{2}, Miguel Daal \orcidicon{0000-0002-1134-2116}\IEEEauthorrefmark{1}}

    \IEEEauthorblockA{\IEEEauthorrefmark{1}University
of California Santa Barbara, Department of Physics and Astronomy, Santa Barbara, CA 93106 USA (email: jennifer\_smith@ucsb.edu)}
    \IEEEauthorblockA{\IEEEauthorrefmark{2}Maybell Quantum Industries, Denver, CO 80221 USA}
}

%+++++++++++++++++++++++++++++++++++++++++++++++++++

% avoiding spaces at the end of the author lines is not a problem with
% conference papers because we don't use \thanks or \IEEEmembership

%\specialpapernotice{(Invited Paper)}

\maketitle

\begin{abstract}
Superconducting arrays often require specialized, high-density cryogenic cabling capable of transporting electrical signals across temperature stages with minimal loss, crosstalk, and thermal conductivity. We report improvements to the design and fabrication of previously published superconducting 53 wt\% Nb-47 wt\% Ti (Nb47Ti) FLexible coAXial ribbon cables (FLAX). We used 3D electromagnetic simulations to inform design changes to improve the characteristic impedance of the cable and the connector transition. We increased the center conductor diameter from 0.003 inches to 0.005 inches which lowered the cable characteristic impedance from $\sim$60 $\bm{\Omega}$ to $\sim$53 $\bm{\Omega}$. This change had a negligible impact on the computed heat load which we estimate to be 5 nW per trace from 1 K to 100 mK with a 1-ft cable. This is approximately half the calculated heat load for the smallest commercially available superconducting coax. We also modified the transition board to include a capacitive coupling between the upper ground plane and signal traces that mitigates the inductive transition. We tested these changes in a 5-trace, 1-ft long cable at 4 K and found the microwave transmission improved from 6 dB to 1.5 dB of attenuation at 8 GHz. This loss is comparable to commercial superconducting coax and 3$\times$ lower than commercial NbTi-on-polyimide flex cables at 8 GHz. The nearest-neighbor forward crosstalk remained less than -40 dB at 8 GHz. We compare key performance metrics with commercially available superconducting coax and NbTi-on-polyimide flex cables and we share initial progress on commercialization of this technology by Maybell Quantum Industries.
\end{abstract}
\IEEEoverridecommandlockouts
\begin{keywords}
Superconducting cables, Superconducting microwave devices, Superconducting filaments and wires, Superconducting resonators.
\end{keywords}

\IEEEpeerreviewmaketitle

\section{Introduction}
% no \PARstart
\IEEEPARstart{S}{uperconducting} devices are driving technology advancements in quantum computing\cite{arute_quantum_2019, acharya_suppressing_2023}, single-photon-counting sensors\cite{walter_mkid_2020, zobrist_wide-band_2019}, low-noise amplifiers\cite{white_readout_2023}, and more. These technologies all require specialized cryogenic wiring capable of transporting electrical signals across temperature stages with minimal loss, crosstalk, and thermal conductivity. The development of high-density cryogenic wiring has proved to be a bottleneck in scaling up superconducting systems that need potentially thousands of cryogenic I/O channels to meaningfully operate \cite{noauthor_ibm_nodate}. 

Progress has been slowed in part by stringent performance requirements that limit wiring solutions to materials that are difficult to work with. Niobium-titanium (NbTi) is one such material that has very low thermal conductivity and is superconducting under 9 K, providing excellent cryogenic transmission and isolation with a fraction of the heat load afforded by other metals\cite{daal_properties_2019}. This unique combination of electrical and thermal properties has led NbTi to become the conductor of choice for cryogenic wiring despite the fact it cannot be soldered and is otherwise difficult to work with and connectorize \cite{noauthor_working_nodate}. Similarly, Teflon-type materials make ideal dielectric mediums because they are flexible with low thermal conductivity and dielectric loss. Unfortunately, these materials come with a host of manufacturing challenges including extremely high melting temperatures, non-trivial phase changes, potential to produce highly-toxic byproducts, and inability to adhere well to other materials \cite{noauthor_cowie_nodate, noauthor_teflon_nodate}.

Despite the fabrication challenges, vendors have managed to fabricate individual cryogenic coax lines by extruding NbTi tubes threaded with Teflon-coated NbTi wire \cite{noauthor_cryogenic_nodate, noauthor_superconducting_nodate, noauthor_sc-08650-nbti-nbti_nodate}. These cables boast excellent transmission and isolation but are semi-rigid and cumbersome to use in small cryogenic volumes. Furthermore, the extrusion process restricts manufacturing to relatively large cross sections resulting excessive heat loads. More recently, processes have been developed that involve sputtering NbTi onto a polyimide material such as Kapton to form stripline traces\cite{walter_laminated_2018, hernandez_microwave_2017, noauthor_crioflex_nodate}. NbTi-on-polyimide cables are more flexible and have lower heat load than the aforementioned semi-rigid cables but have higher dielectric loss that scales with frequency. This loss can be prohibitive to applications carrying RF or Microwave signals from cryogenic devices before amplification where the signals are small and vulnerable to noise. Neither of these leading commercial options met the transmission, crosstalk, and heat load requirements for our application delivering 4-8 GHz signals to an array of 20,000+ microwave kinetic inductance detectors (MKIDs) at 10 mK \cite{meeker_design_2015, walter_mkid_2020}. 

Unable to buy suitable cryogenic wiring, we previously developed our own in-house cable manufacturing process at the University of California, Santa Barbara that involves spot welding NbTi foil around PFA-coated NbTi wires to form a flexible coaxial ribbon cable (FLAX) \cite{smith_flexible_2021}. This technology aims to replicate the near-perfect isolation and transmission offered by semi-rigid, superconducting coax but in a smaller, more flexible format that reduces heat load and allows for increased I/O density. In \cite{smith_flexible_2021}, we demonstrated competitive heat load and isolation but the transmission loss could prove to be a dominant noise source for highly-sensitive experiments like those requiring quantum-noise-limited amplifiers. Another drawback was the highly variable, manual, and labor intensive manufacturing process \cite{smith_flexible_2021}.

In this paper, we present improvements to the cable design and show 3x improved cable transmission without negatively impacting the isolation or heat load. We share initial progress by Maybell Quantum Industries on a roll-to-roll manufacturing system to streamline and standardize cable fabrication in preparation for commercialization. Lastly, we compare the latest performance data with the smallest commercially available semi-rigid superconducting coax from CryoCoax\footnote{CryoCoax - Intelliconnect, 448 Old Lantana Road, Crossville, TN, USA. \vskip1pt \hskip5ptP/N: BCB012}\cite{noauthor_cryogenic_nodate} and the NbTi-on-polyimide Cri/oFlex\textsuperscript \textregistered  3 from Delft Circuits\footnote{Delft Circuits, 2627 AN, Delft, Netherlands. \vskip1pt \hskip5ptP/N: Cri/oFlex\textsuperscript \textregistered  3}\cite{noauthor_crioflex_nodate}.

\section{Manufacturing Improvements}

\begin{figure}
\includegraphics[width=3.4in]{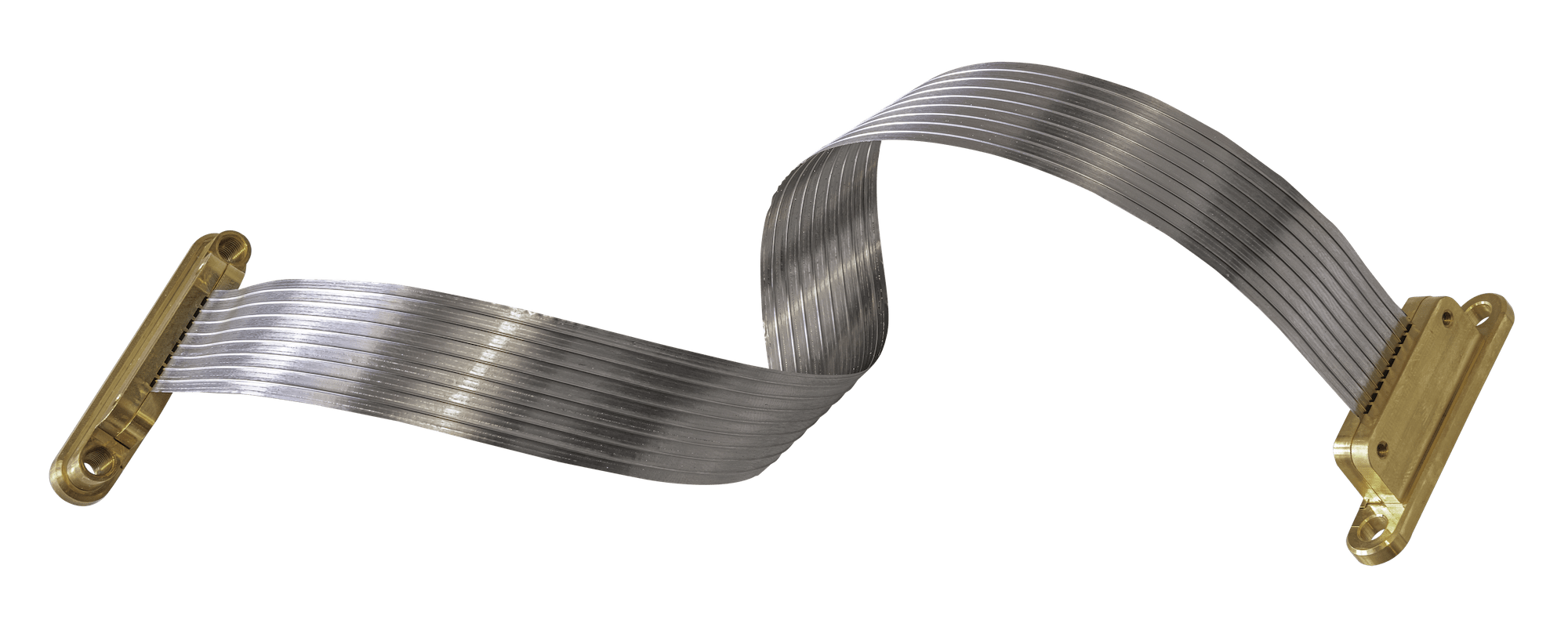}
\caption{Eight-trace Maybell cable prototype manufactured using linearly actuated mechanical stage with laser welding. Two sheets of 1 mil thick NbTi foil are welded around PFA-coated NbTi wires with a continuous laser weld.}\label{fig:picture}
\end{figure}

The previous iteration of the FLAX cable was manufactured using individual, hand-placed spot welds to act as vias between the cable traces \cite{smith_flexible_2021}. While effective, this process was slow and error-prone because any welds placed too close to the center wire result in an electrical short, rendering the cable useless. Maybell has developed a laser welding cable manufacturing system and is developing a roll-to-roll apparatus that will greatly simplify and speed up speed up cable manufacturing as well as increase overall uniformity. A prototype sample laser-welded using a linearly actuated mechanical stage is shown in Fig.~\ref{fig:picture}. A close up of the continuous laser welds surrounding two wires is show in Fig.~\ref{fig:manufacturing}. We expect cables manufactured with this system to show enhanced durability and better impedance control due to the precise, continuous laser welds. We also note the proposed roll-to-roll process allows cables to be made with arbitrary length, giving more flexibility to cryogenic routing and design.

\section{Design Improvements}

Two main design improvements were made to correct for impedance mismatches in the cable and improve the microwave transmission. First, the characteristic impedance of the cable was adjusted closer to 50 $\Omega$. Second, the connectorization was modified to reduce reactive impedance reflections off the connector ends.

\subsection{Characteristic Impedance}
\begin{figure}
\includegraphics[width=3.4in]{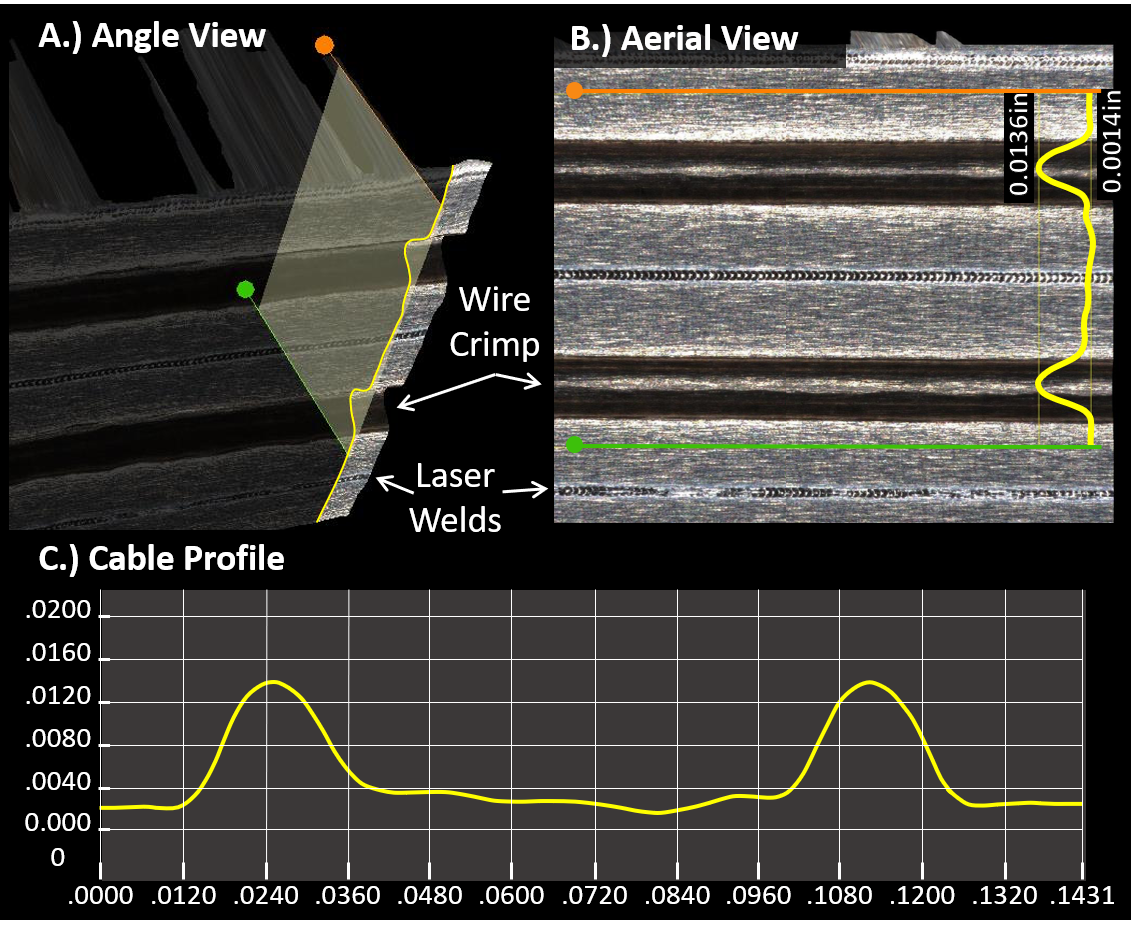}
\caption{Manufacturing data obtained from JT Automation during a prototype cable manufacturing run. A.) Angle view showing a single sheet of NbTi foil with two crimps made to house wires with continuous laser welds in-between. B.) Aerial view of the NbTi foil showing the same welds and crimps pictured to the left. C.) Cable profile measured in inches. In all views the cable profile is highlighted in yellow.}\label{fig:manufacturing}
\end{figure}

The characteristic impedance of the cable was originally designed in \cite{smith_flexible_2021} assuming a perfect circular cross section for the inner and outer conductor geometry. In reality, the manufacturing process produces an outer conductor profile more parabolic than semicircular (see Fig.~\ref{fig:manufacturing}). This results in the cable cross section appearing more elliptical (see Fig.~\ref{fig:exploded}) which causes the outer conductor to be effectively further from the inner-conductor, elevating the cable characteristic impedance.

We modeled the actual coaxial shape as a spline in Ansys HFSS and determined increasing the center conductor diameter from 0.003 inches to 0.005 inches (without changing the dielectric thickness) would effectively bring the inner and outer conductors closer together and bring the impedance back down to the target 50 $\Omega$.

\subsection{End Inductance}

\begin{figure}
\includegraphics[width=3.5in]{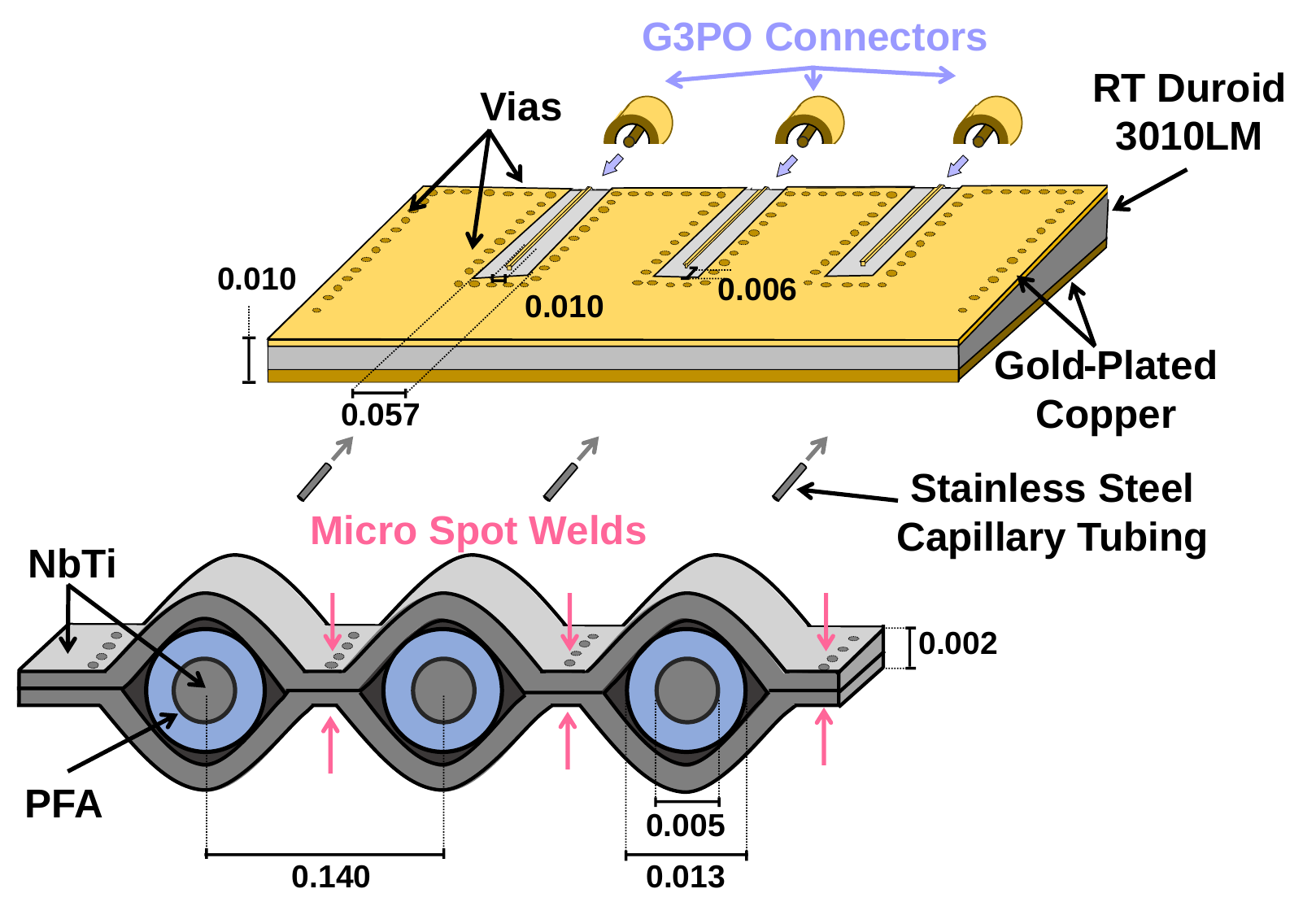}
\caption{Exploded view of cable-end assembly with key dimensions shown in inches. Drawing is not to scale. From top/back to bottom/front:  G3PO half-shell connectors are soldered to the transition board. The transition board features 50 $\Omega$ grounded coplanar waveguide structures with the upper ground planes connected in front to create capacitance between the upper ground planes and signal traces. The FLAX cable center conductors are crimped into stainless steel capillary tubes and soldered to the center traces. The FLAX ground shield is spot welded to the upper ground planes. The cable cross section shows the PFA (blue) insulated NbTi (grey) wire set in parabolic crimps made in the shared NbTi foil ground shield. The two sides of the shield are mechanically and electrically bonded with micro spot welds less than $\lambda/16\simeq$ 2 mm (at 8 GHz) apart that run in-between the traces down the length of the cable. Figure is modified from \cite{smith_flexible_2021}.}\label{fig:exploded}
\end{figure}

When traveling out of the cable and into the connector, the electric field transitions from a coaxial mode in the cable to a coplanar waveguide mode in the transition board, then back to a coaxial mode in the connector. The center conductors, which behave inductively, are relatively continuous in the transition; however, the surrounding ground planes, which provide a balancing capacitance, undergo a large geometry shift from 3D to 2D and back. The authors theorize that the disruption to the 3D ground plane reduces capacitance in the transition and causes the connector ends to behave inductively. This can be seen in the transmission spectrum from \cite{smith_flexible_2021}, which shows ripples in the transmission consistent with the fundamental mode of the cable growing with increasing frequency. This is consistent with reflections off inductive features in the cable ends because the impedance of a perfect inductor grows linearly with frequency, i.e., $Z_L = j\omega L$. 

\begin{figure}
\includegraphics[width=3.4in]{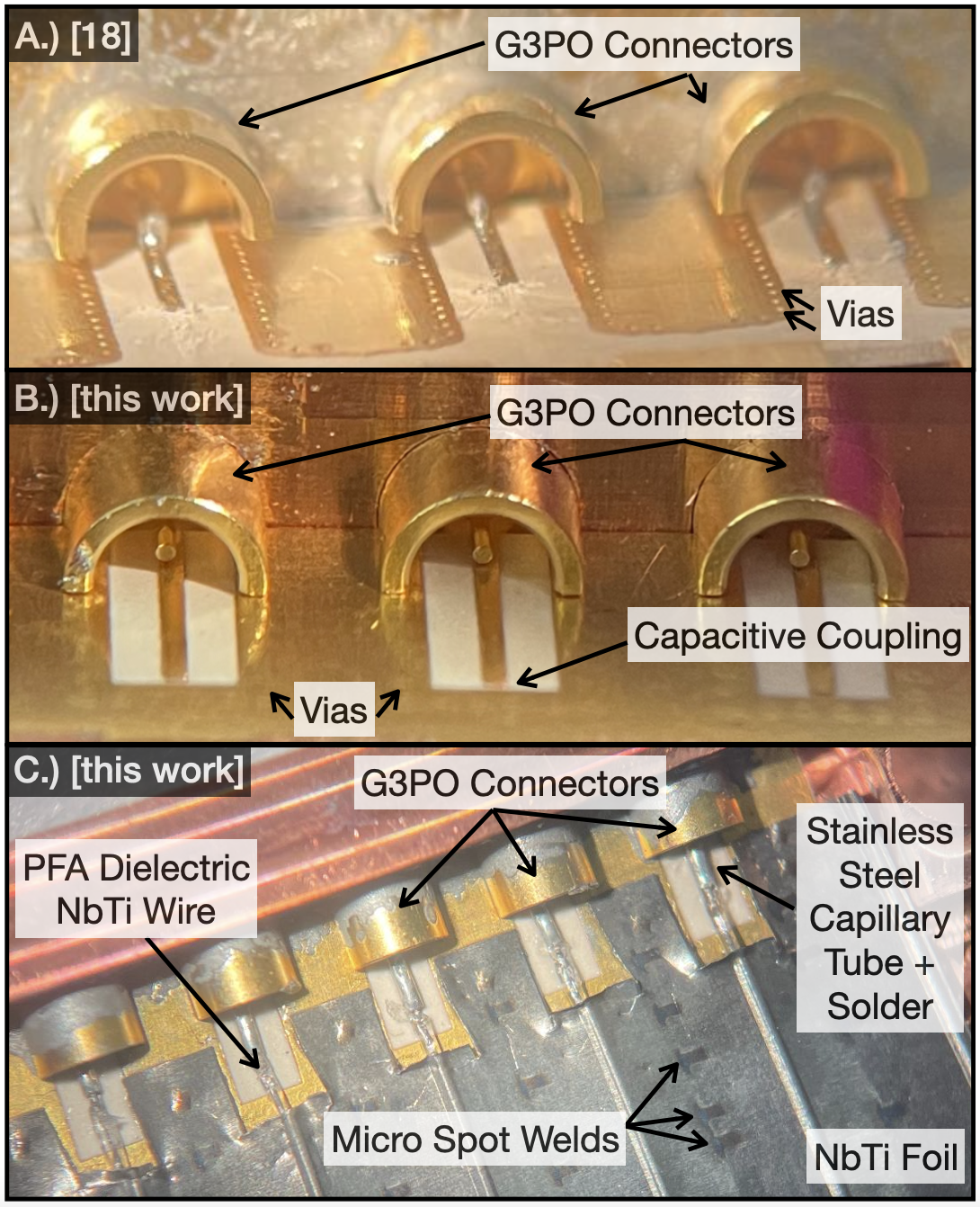}
\caption{A.) Box end housing and grounded coplanar waveguide transition board design used in \cite{smith_flexible_2021} that yielded excess end inductance. B.) Box end and modified grounded coplanar waveguide design with capacitive coupling used in this work. C.) Connectorization scheme showing how cable is connectorized in this work. The cable ground is spot welded to the transition board ground. The cable center wire is exposed by cutting the foil ground and then crimping a stainless steel tube onto the wire. The tube is soldered to the transition board. The transition board and G3PO connectors are soldered into the box housing.}\label{fig:connectorization}
\end{figure}

The authors used Ansys HFSS to simulate the connector transition and determined the inductance in the cable ends could be mitigated by including a capacitive feature in the transition board. This was accomplished by modifying the upper coplanar waveguide ground planes on the transition board, bringing them together in front of the signal trace with a 0.006 inch gap (see Fig.~\ref{fig:exploded} and Fig.~\ref{fig:connectorization} ).

\section{Performance Characterization}
\begin{table*}[!htbp] 
\caption{ Summary of thermal, mechanical, and microwave properties of superconducting coaxial ribbon cable, commercial semi-rigid superconducting coax, and  commercial superconducting polyimide flex cables}\label{table:properties}
\vskip2pt
\centerline{
\vbox{\offinterlineskip
\hrule
\halign{&\vrule#&
\strut\quad#\hfil\quad\cr
&\strut&&\multispan3\hfil {\bf Thermal Load\footnote{} }\hfil&&\multispan9\hfil {\bf Mechanical}\hfil &&\multispan3\hfil {\bf Microwave}\footnote{}\hfil &\cr
&{\bf Cable}&&\multispan3\hfil per trace [nW] \hfil&&\multispan9\hfil All Dimensions [inches] \hfil && \multispan3\hfil Values at 8 GHz \hfil &\cr
& &&\multispan3\hfil 100mK to \hfil&& Connector && OD && Min. Inside && Conductor && Dielectric && Crosstalk  \hfil &&
Attenuation &\cr
&\omit && 1 K & &4 K && Pitch && (${\diameter}$) && Bend Radius && Material && Material && [dB] && [dB] &\cr
height2pt&\omit&&\omit&&\omit&&\omit&&\omit&&\omit&&\omit&&\omit&\cr
\noalign{\hrule}
height2pt&\omit&&\omit&&\omit&&\omit&&\omit&&\omit&&\omit&&\omit&\cr
& FLAX [This Work] && $5$ && $245$ && \hskip9pt$0.140$ && $0.02$ && $0.08$ && NbTi && PFA && \hskip9pt$-40$ && $1.5$ &\cr
& FLAX \cite{smith_flexible_2021} && $5$ && $238$ && \hskip9pt$0.140$ && $0.02$ && $0.08$ && NbTi && PFA && \hskip9pt$-40$ && $7$ &\cr
& Semi-Rigid\footnote{} && $7$ && $370$ && $>0.5$ && $0.90$0 && $0.13$ && NbTi && PTFE && $<-60$ && $0.5$ &\cr
& Polyimide-Flex\footnote{} && TBD && TBD && \hskip9pt$0.3$ && $0.01$\footnote{} && $0.04$ && NbTi && Polyimide && \hskip9pt $-60$ && $7$ &\cr}
\hrule}}
\hskip20pt \footnotesize{$^3$ Computed using a cable length of 1 foot. } 
\vskip1pt \hskip20pt \footnotesize{$^4$ Values reported for $\sim$1-ft long cables.} 
\vskip1pt \hskip20pt \footnotesize{$^{5}$ CryoCoax 0.034 inch NbTi/NbTi, P/N: BCB012 \cite{noauthor_cryogenic_nodate}.}
\vskip1pt \hskip20pt \footnotesize{$^{6}$ Delft Circuits Cri/oFlex\textsuperscript \textregistered 3 with NbTi conductor \cite{noauthor_crioflex_nodate}.}
\vskip1pt \hskip20pt \footnotesize{$^{7}$ For the microstrip geometry this is the total cable thickness.}

\end{table*}

Design modifications were tested using a 5-trace, 1-ft long cable manufactured in-house at the University of California, Santa Barbara using the same process described in \cite{smith_flexible_2021}. We compare our results with the previous iteration of the cable as well as the smallest commercially available superconducting coax from CryoCoax (P/N: BCB102)\cite{noauthor_cryogenic_nodate} and the only commercially available NbTi-on-polyimide flex cable from Delft Circuits (P/N: Cri/oFlex\textsuperscript \textregistered  3) \cite{noauthor_crioflex_nodate}. Key thermal, mechanical, and electrical properties are summarized in Table ~\ref{table:properties}.

\begin{figure}
\includegraphics[width=3.4in]{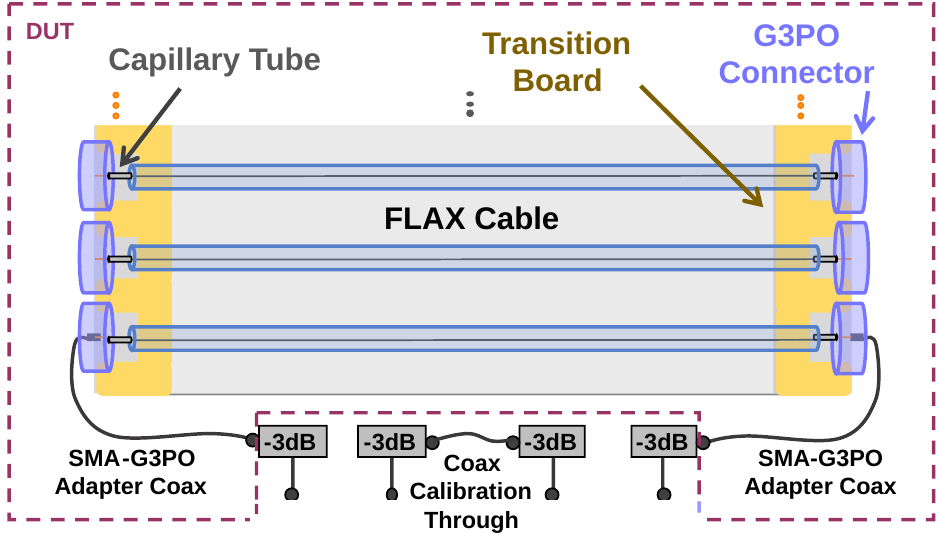}
\caption{Schematic diagram depicting the device under test  circuit at 4 K. Figure is modified from \cite{smith_flexible_2021} }\label{fig:DUT}
\end{figure}

Transmission loss ($S_{21}$), crosstalk ($S_{41}$), and time domain reflectometry were measured at 4 K with a network analyzer. The device under test circuit consisted of the assembled FLAX cable with 3 dB cryo-attenuators obtained from XMA\footnote{XMA Corporation-Omni Spectra, 7 Perimeter Road, Manchester, NH. \vskip1pt \hskip4pt P/N: 2082-6040-03-CRYO} and 25 cm nonmagnetic SMA-to-G3PO adapter coaxial cables obtained from Koaxis\footnote{Koaxis RF Cable Assemblies, 2081 Lucon Road, Schwenksville, PA. \vskip1pt \hskip4pt P/N: AO10-CC047C-YO18} on either end (see Fig.~\ref{fig:DUT}). A short commercial coax through line was used as a calibration reference for fridge wiring as in \cite{smith_flexible_2021}.  

\subsection{Transmission}

\begin{figure*}
\includegraphics[width=\textwidth]{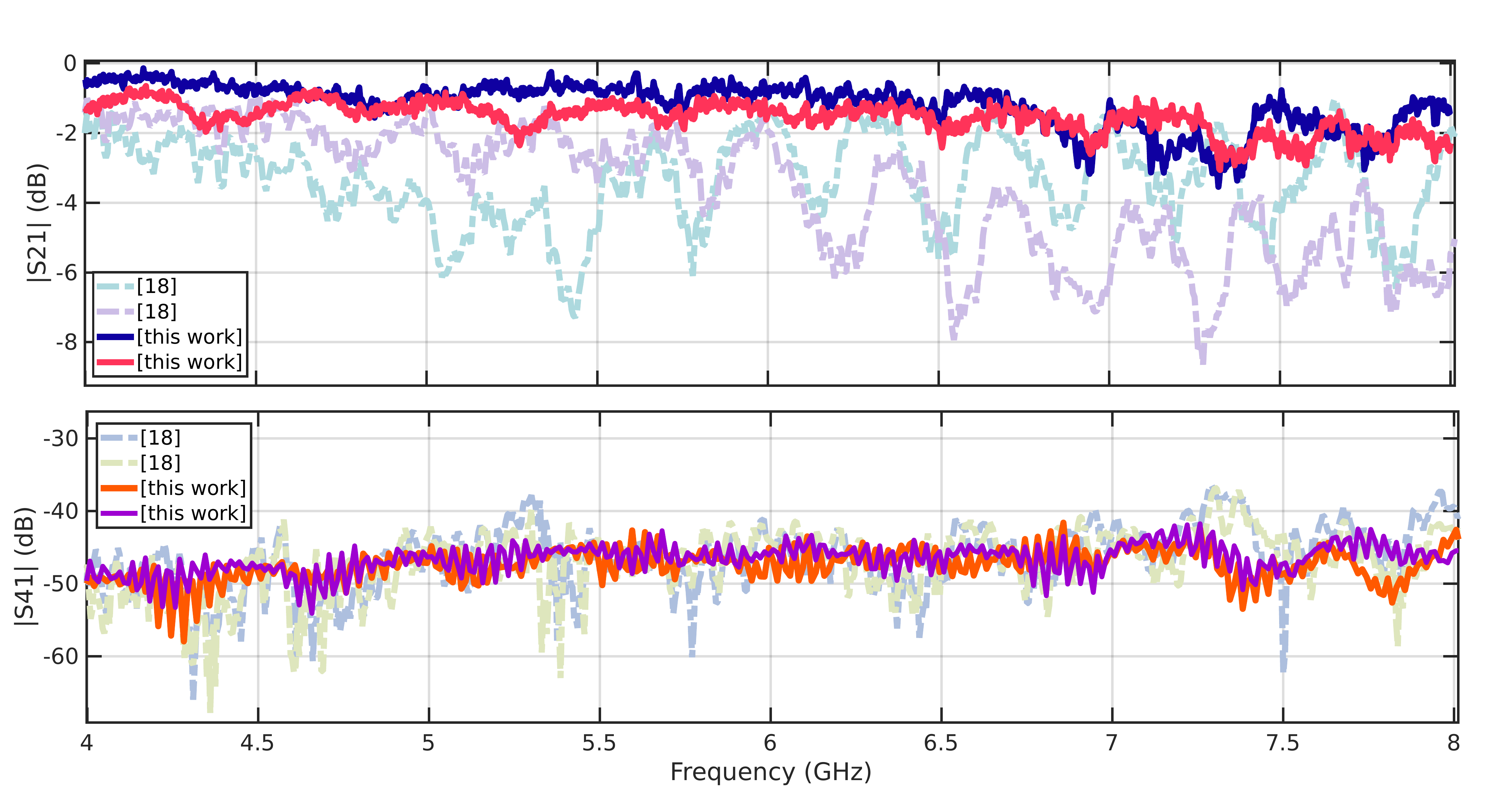}
\caption{Top: $S_{21}$ (transmission) measurement of sample traces from \cite{smith_flexible_2021} (dashed) and this work (solid) at 4 K. Bottom: $S_{41}$ (nearest neighbor far-end crosstalk) measurement of sample FLAX traces from the same cables at 4 K.}\label{fig:transmission}
\end{figure*}

Transmission measurements ($S_{21}$) at 4 K for two traces from the same test cable are plotted in Fig.~\ref{fig:transmission} along with sample traces from the previous iteration of the cable first published in \cite{smith_flexible_2021}. The transmission is improved from at most 7 dB of loss in \cite{smith_flexible_2021} to 1.5 dB of loss at 8 GHz in this work. The loss is now on-par with commercial semi-rigid superconducting coax cables and 3x better than commercial NbTi-on-polyimide cables (see Table ~\ref{table:properties}).

Previously, the authors had hypothesized ripples in the transmission spectrum were caused by inductance in the connector transition that produced reflections off the cable ends \cite{smith_flexible_2021}. The transmission curves from this work do not show the same ripples increasing with frequency as the previous iteration, suggesting the capacitive tuning in the transition board successfully mitigated the inductive connectorization. As expected, we find that connectorization plays a critical role in the overall cable performance.

\begin{figure}
\includegraphics[width=3.4in]{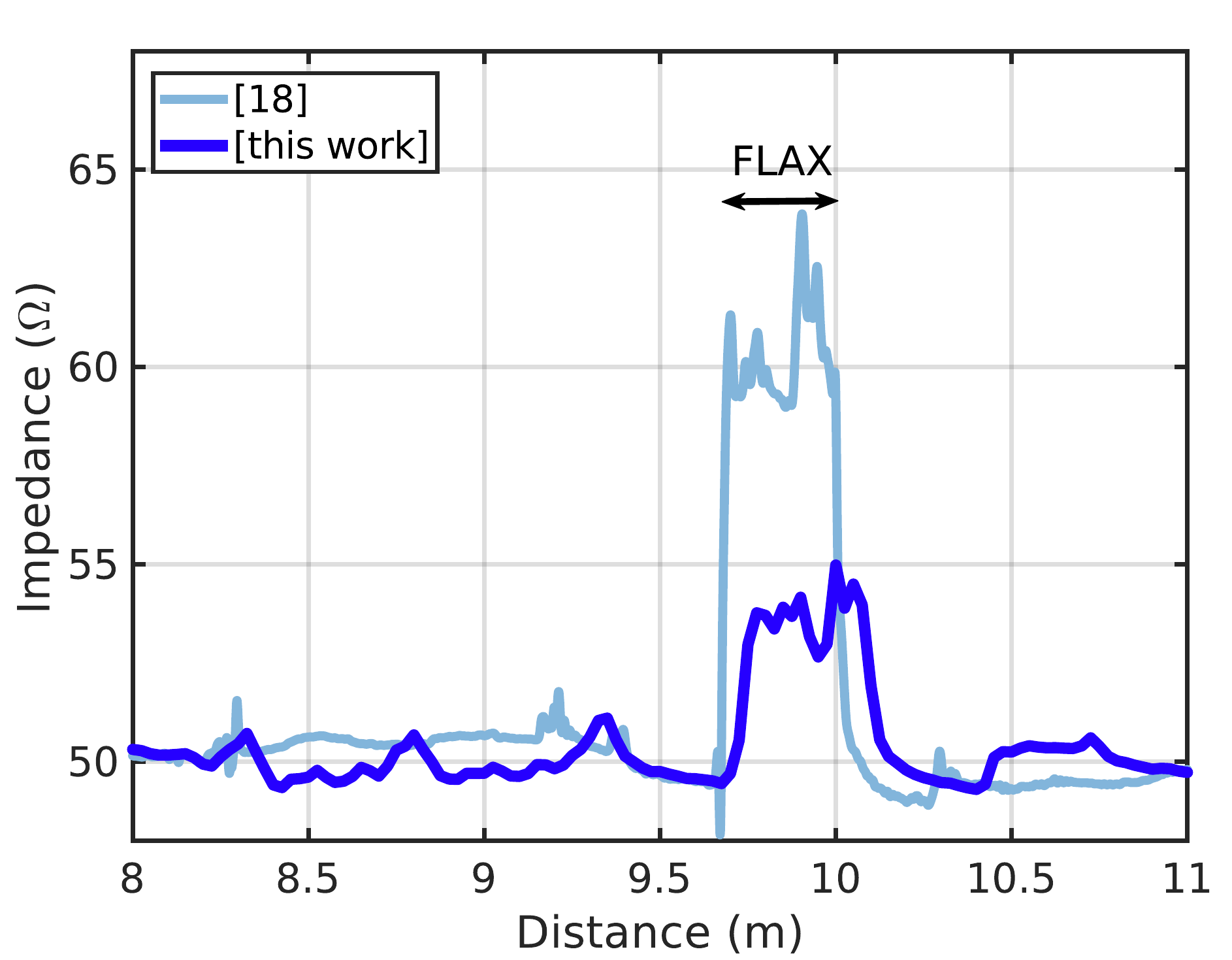}
\caption{Time Domain Reflectometry (TDR) measurement of the cryogenic signal path showing the characteristic impedance at lengths along the signal path at 4 K. Traces are shown from \cite{smith_flexible_2021} (light blue) and this work (dark blue). Commercially available 50 $\Omega$ coaxial cables border the FLAX cables highlighted by the double arrow. Note the 4 K TDR measurement is accurate to $\pm$ 3 $\Omega$. }\label{fig:TDR}
\end{figure}

An updated time domain reflectometry measurement is shown in Fig.~\ref{fig:TDR} and confirms the larger center conductor wire lowered the impedance from $\sim$60 $\Omega$ to $\sim$53 $\Omega$. The impedance being closer to 50 $\Omega$ is likely contributing to the improved transmission at lower frequencies where the insertion loss dominates over the inductive ripple. The improved characteristic impedance and transmission spectra are consistent with 3D electromagnetic simulations the authors used to model design modifications to the center wire and transition board.

\subsection{Crosstalk}
The authors measured both the near-end crosstalk ($S_{31}$) and the far-end crosstalk ($S_{41}$) and found the crosstalk levels to be the same. This suggests all crosstalk is happening in the transition board and connector ends. This is consistent with the flexible coaxial ribbon cable providing a 3D superconducting ground around each wire making crosstalk in the cable itself virtually impossible. Only ($S_{41}$) is shown in Fig.~\ref{fig:transmission} for comparison with \cite{smith_flexible_2021}. We find the crosstalk remained less than -40 dB which is competitive with commercial semi-rigid and Kapton-flex cables. We hypothesize the crosstalk may be further reduced by adding 3D shielding on the surface of the transition board.

\subsection{Thermal Conductivity}
\begin{figure}
\includegraphics[width=3.4in]{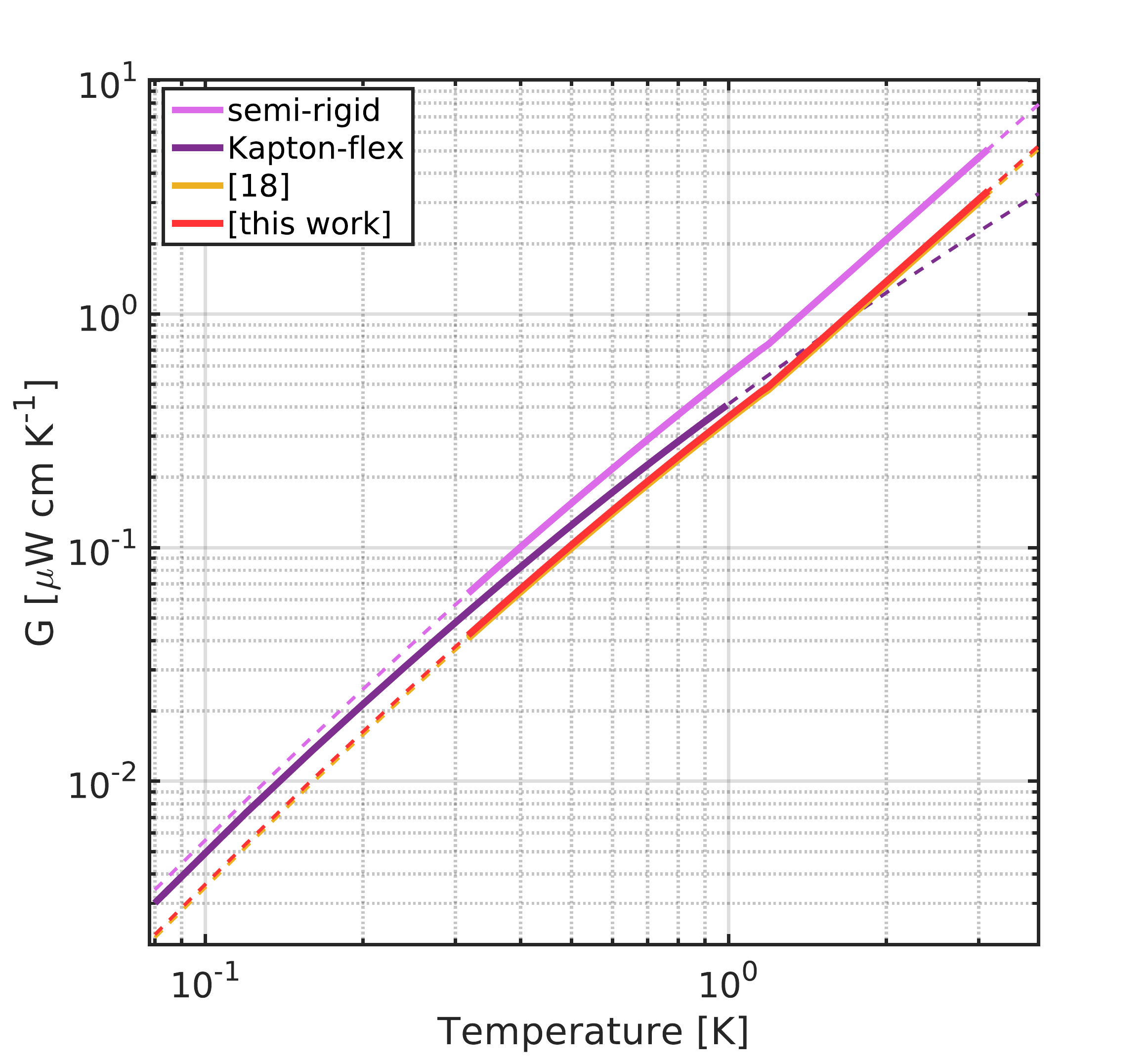}
\caption{We computed a cable thermal conductivity $G(T)$ in units of $\mu \mathrm{W cm K}^{-1}$ by summing the thermal conductivity of each constituent material weighted by the cross section\cite{kushino_thermal_2005, walter_laminated_2018, smith_flexible_2021}. The smallest commercially available superconducting coax from CryoCoax (P/N: BCB012) labeled ``semi-rigid" (pink) is compared with this work (salmon) and \cite{smith_flexible_2021} (gold). The authors were unable to obtain cross-sectional dimensions for the Delft Cri/oFlex\textsuperscript \textregistered  3 cable and so instead display a fictitious NbTi/Kapton cable, labeled ``Kapton-flex" (purple), where the Kapton and NbTi materials have been assigned the same cross-sectional areas as the dielectric and conductor in this work, respectively. The differences in heat load between the semi-rigid (pink), \cite{smith_flexible_2021} (gold), and this work (salmon) are due purely to geometrical differences whereas the difference between the Kapton-flex (purple) and this work (salmon) is due purely to difference in dielectric materials. Solid lines are computed using literature values for Nb47Ti\cite{daal_properties_2019}, PTFE\cite{kushino_thermal_2005}, and Kapton \cite{walter_laminated_2018, kellaris_sub-kelvin_2014}. PTFE values were used to estimate the PFA dielectric in the FLAX cable. Dashed lines indicate extrapolation. Figure is modified from \cite{smith_flexible_2021}.}\label{fig:heatload}
\end{figure}

Following previous convention, we compute $G(T)$, a length-dependent cable thermal conductivity, by summing literature values of cable materials weighted by their cross sections (see Fig.~\ref{fig:heatload}) \cite{kushino_thermal_2005, smith_flexible_2021, walter_laminated_2018}. The heat load between temperature stages can be computed by integrating values in Fig.~\ref{fig:heatload} from $T_1$ to $T_2$ ($T_1<T_2$) and dividing by the cable length. Results of this calculation are presented for select temperatures and cables in Table ~\ref{table:properties}. $G(T)$ is plotted for the superconducting coaxial ribbon cable in this work, the previous iteration \cite{smith_flexible_2021}, and the smallest commercially available superconducting coax from CryoCoax (P/N: BCB012)\cite{noauthor_cryogenic_nodate} in Fig.~\ref{fig:heatload}. Delft Circuits declined request to share cross-sectional area information for the superconducting polyimide Cri/oFlex\textsuperscript \textregistered  3 cable (Delft Circuits, personal communication, May 2023) and so we instead show a hypothetical NbTi-on-polyimide cable where the polyimide\footnote{approximated as the Nikaflex DuPont/Nikkan Kapton film\cite{noauthor_dupont_nodate} used in \cite{walter_laminated_2018}.} cable dielectric and conductor have the same cross sectional areas as the dielectric and conductor in this work, respectively. In this way, the differences in heat load between the three superconducting coaxes in Fig.~\ref{fig:heatload} are purely geometrical whereas the difference between this work and the polyimide/Kapton is due solely to materials differences in the dielectric.

We find the heat load increased slightly between \cite{smith_flexible_2021} and this work due to the increased center conductor diameter (see Table ~\ref{table:properties}). Despite the increase, the cable presented in this work remains roughly half as thermally conductive as the smallest commercially available superconducting coax. While the heat load of the Cri/oFlex\textsuperscript \textregistered  3 cables with NbTi conductor has not been directly measured, Delft Circuits suggests it may be $\sim$10\% of the heat load of the same polyimide-flex cables with Ag conductor, i.e. 6x more thermally conductive from 1 K to 100 mK than the cable presented in this work (Delft Circuits, personal communication, May 2023)\cite{noauthor_crioflex_nodate}. This may be because the polyimide material makes up the bulk of the cross section and is approximately 5 times more thermally conductive than NbTi below 1 K \cite{daal_properties_2019}. In this work, we hypothesize the cable heat load is dominated by the NbTi conductor because \cite{daal_properties_2019} suggests NbTi is approximately ten-times more thermally conductive than PTFE below 1 K. To minimize heat load in cryogenic cables, we suggest NbTi/PTFE/PFA solutions minimize conductor cross section whereas NbTi/Kapton/polyimide technologies minimize dielectric cross section.

% The following statement makes the two columns on the last page more
% or less of equal length.  Placement of this command is by trial and error.
%\vfil\eject

\section{Conclusion}
We have improved the transmission 3x without negatively impacting the heat load or isolation and made progress on commercial manufacturing of the superconducting flexible coaxial ribbon cable presented in \cite{smith_flexible_2021}. We adjusted the center conductor diameter and re-designed the transition board to improve the transmission loss while retaining strong signal isolation and low heat load.

The cable technology presented in this work has been deployed in two superconducting-array-based instruments: the MKID Exoplanet Camera (MEC)\cite{walter_mkid_2020} and the MagAO-X MKID Instrument (XKID)\cite{swimmer_mkid_2022}. In both cases, FLAX cables replaced our group's previously developed laminated NbTi-on-Kapton microstrip cables \cite{walter_laminated_2018}. In MEC, the improvement in signal integrity increased the amount of usable superconducting detectors in the array by $\sim$20\%. In XKID, the average MKID energy resolution--- a key superconducting detector performance metric, increased by a factor of 4.

This cable technology has already demonstrated promising performance gains in superconducting detector array instruments. With the modifications in this work along with future improvements and commercialization by Maybell Quantum Industries, we see the potential for this technology to be widely applicable to superconducting device systems. Future improvements may include adding 3D shielding in the transition board which would allow for higher connector pitch density, lowering the heat load per trace and potentially improving the crosstalk. 

Overall, we find this cable technology to be superior to commercial options for our applications building high-density superconducting detector arrays. The high electrical and thermal isolation coupled with low loss and connectorized with high-density, coaxial push-on connectors makes for an easy-to-use cryogenic wiring solution. We are looking forward to future improvements led by Maybell Quantum Industries and we expect these results will be especially promising for groups looking to operate high-density, superconducting microwave device arrays.

\section*{Acknowledgment}
J. P. Smith is supported by a NASA Space Technology Research Fellowship under grant number 80NSSC19K1126. This work was supported by a DARPA SBIR Grant under grant number W912CG23C0002.
\bibliographystyle{IEEEtran}
\bibliography{references}
\end{document}